\documentclass[a4paper]{article}
\usepackage{spconf,amsmath,graphicx,cite}
\usepackage[vlined,linesnumbered,ruled,resetcount]{algorithm2e}
\usepackage{epstopdf}

\let\oldnl\nl
\newcommand{\nonl}{\renewcommand{\nl}{\let\nl\oldnl}}

\newcommand{\nextnr}{\stepcounter{AlgoLine}\ShowLn}

\DeclareMathOperator*{\argmax}{\arg\!\max}

\title{BAYESIAN CALIBRATION USING DIFFERENT PRIOR DISTRIBUTIONS: AN ITERATIVE MAXIMUM A POSTERIORI APPROACH FOR RADIO INTERFEROMETERS}
\name{V. Ollier$^*$ $^\dagger$,  M. N. El Korso$^\ddagger$, A. Ferrari$^\mathsection$, R. Boyer$^\dagger$ and P. Larzabal$^*$\thanks{This work was supported by the
following projects: MAGELLAN (ANR-14-CE23-0004-01), ON FIRE project (Jeunes Chercheurs GDR-ISIS) and ANR ASTRID project MARGARITA (ANR-17-ASTR-0015).}}
\address{%
    $^*$ SATIE, UMR 8029, ENS Paris-Saclay, Cachan, France\\
    $^\dagger$ L2S, UMR 8506, Universit\'{e} Paris-Sud, Gif-sur-Yvette, France\\
    $^\ddagger$ LEME, EA 4416, Universit\'{e} Paris-Nanterre, Ville d'Avray, France \\
    $^\mathsection$ Laboratoire J.L. Lagrange, UMR 7293, Universit\'{e} Nice Sophia-Antipolis, France\\
}
\begin{document}

\maketitle
\begin{abstract}

In this paper, we aim to design robust estimation techniques based on the compound-Gaussian (CG) process and adapted for calibration of radio interferometers. The motivation beyond this is due to the presence of outliers leading to an unrealistic traditional Gaussian noise assumption.
Consequently, to achieve robustness, we adopt a maximum a posteriori (MAP) approach which exploits Bayesian statistics and follows a sequential updating procedure here. The proposed algorithm is applied in a multi-frequency scenario in order to enhance the estimation and correction of perturbation effects. Numerical simulations assess the performance of the proposed algorithm for different noise models, Student's t, K, Laplace, Cauchy and inverse-Gaussian compound-Gaussian distributions \textit{w.r.t.} the classical non-robust Gaussian noise assumption.

%In order to assess the robustness of the proposed statistical procedure \textit{w.r.t.} outliers, comparisons are conducted between different prior distributions, leading to various heavy-tailed noise models.

\end{abstract}
\section{Introduction}
\label{sec:intro}

Robust calibration  in radio astronomy amounts to estimating all environmental and instrumental perturbation effects which corrupt the signal of interest in a non-Gaussian environment. Indeed, radio interferometric data are affected by the occurrence of outliers due, \textit{e.g.}, to man-made radio frequency interferences or unmodelled weak sources \cite{raza2002spatial,yatawatta2014robust,zarkaNENUFAR}. Therefore, estimation under a conventional Gaussian noise model does not perform optimally and robust calibration is required. Furthermore, unknown perturbation effects being highly variable in frequency \cite{noordam2010meqtrees}, exploiting their variation \textit{w.r.t.} frequency improves estimation performance in the calibration process while maintaining an acceptable computational cost \cite{yatawatta2015distributed,martinjournalnew}. To this end, calibration can be reformulated as a distributed consensus optimization problem and solved thanks to the alternating direction method of multipliers (ADMM) \cite{boyd2011distributed}.

%van2013lofar
%dewdney2009square
%thompson2008interferometry
%wijnholds2010calibration
%noordam2004lofar
%noordam1996measurement
%hamaker1996understanding
%smirnov2011revisiting
%smirnov2011revisiting2
%lochner2015bayesian
%yatawatta2009radio, kazemi2011radio
%dempster1977maximum, feder1988parameter, mclachlan2007algorithm
%madsen1999methods, nocedal2006numerical, transtrum2012improvements
%kazemi2013robust  -> Student's t-distribution is considered
%jay2002detection, yao2003spherically
%lange1989robust, yatawatta2014robust
%Depending on the texture distribution considered, it is possible to generate various noise models different from the Gaussian one. 
%wang2006maximum -> ML method
%zhang2015maximum

%,rangaswamy1993spherically
To investigate non-Gaussian modeling, we consider a two-scale compound-Gaussian (CG) process, also referred to as a spherically invariant random variable in the radar community \cite{ollila2012complex,jay2002detection}, which is represented as the product of a positive texture term with a given a priori distribution, multiplied by a complex speckle component following a zero-mean Gaussian distribution. Such a model is suitable for practical scenario and widely used in signal processing applications \cite{conte1987characterisation,gini2002vector}. In our scenario, the statistical distribution of the texture parameter is unknown. Thus, we study different prior distributions for the texture, available in closed-form, and generating various non-Gaussian (heavy-tailed) noise models. In order to estimate perturbation effects and noise parameters, we devise an iterative algorithm where optimization is performed successively \textit{w.r.t.} each unknown parameter, while fixing the others. A maximum a posteriori (MAP) approach is adopted to exploit the statistical information and incorporate the texture prior distribution, leading to the so-called iterative maximum a posteriori estimator (IMAPE).

In this paper, we use the following notation: symbols $\left( \cdot \right) ^{T}$, $\left( \cdot \right)^{\ast }$, $\left( \cdot \right) ^{H}$ denote, respectively, the transpose, the complex conjugate and the Hermitian operator. The Kronecker product is represented by $\otimes$, $\mathrm{E}\{\cdot\}$ denotes the expectation operator and $\mathrm{diag}\{\cdot\}$ converts a vector into a diagonal matrix. The trace and determinant operators are, respectively, given by $\mathrm{tr}\left\{ \cdot \right\} $ and $ |\cdot|$. Finally, the symbol $\mathbf{I}_{2}$ represents the $2 \times 2$ identity matrix, $\mathrm{vec}(\cdot)$ stacks the columns of a matrix on top of one another, $j$ is the complex number whose square equals $-1$, $\Gamma(\cdot)$ is the gamma function and $\Psi(\cdot)$ the digamma function, such that, $\Psi(x)=\partial \ln \left(\Gamma(x)\right) / \partial x$.

\section{Data model}
\label{sec:model}

\subsection{Direction dependent regime}

In radio astronomy, propagation of the \textit{i}-th $2$-dimensional incoming signal $\mathbf{s}^{[f]}_i$ at frequency $f \in \mathcal{F}=\{f_1,\hdots,f_F\}$ generates the following voltage at antenna \textit{p} \cite{KerThesis2012}
\begin{equation}
\label{start}
\mathbf{v}^{[f]}_{i,p}(\boldsymbol{\theta}^{[f]})=\mathbf{J}^{[f]}_{i,p}(\boldsymbol{\theta}^{[f]})\mathbf{s}^{[f]}_i
\end{equation}
where the so-called $2 \times 2$ Jones matrix $\mathbf{J}^{[f]}_{i,p}(\boldsymbol{\theta}^{[f]})$ is parametrized by unknown vector $\boldsymbol{\theta}^{[f]}$ and accounts for all perturbation effects along the propagation path \cite{thompson2008interferometry}.

For each antenna pair $(p,q)$, the interferometer measures cross-correlations between voltages, given by
\begin{align}
\label{ME}
\nonumber
\mathbf{S}^{[f]}_{pq}(\boldsymbol{\theta}^{[f]})& =\mathrm{E}\left\{\left(\sum_{i=1}^{D}\mathbf{v}^{[f]}_{i,p}(\boldsymbol{\theta}^{[f]})\right)\left(\sum_{i=1}^{D}\mathbf{v}_{i,q}^{[f]^H}(\boldsymbol{\theta}^{[f]})\right)\right\} \\ & =
\sum_{i=1}^{D}\mathbf{J}^{[f]}_{i,p}(\boldsymbol{\theta}^{[f]})\mathbf{C}^{[f]}_{i}
\mathbf{J}_{i,q}^{[f]^H}(\boldsymbol{\theta}^{[f]})
\end{align}
where $D$ is the number of known calibrator sources, $(p,q)\in \{1, \ldots, M\}^2$ with $M$ the number of sensor elements and $\mathbf{C}^{[f]}_{i}=\mathrm{E}\{\mathbf{s}^{[f]}_{i}\mathbf{s}^{[f]^H}_{i}\}$ stands for the known source coherency matrix \cite{smirnov2011revisiting}. 
We rewrite (\ref{ME}) as a $4 \times 1$ measurement vector and add the noise contribution, such that, 
\begin{equation}
\label{express_xpq}
\mathbf{x}^{[f]}_{pq}=
\sum_{i=1}^{D}\mathbf{s}^{[f]}_{i,pq}(\boldsymbol{\theta}^{[f]})+\mathbf{n}^{[f]}_{pq}
\end{equation}
in which $\mathbf{s}^{[f]}_{i,pq}(\boldsymbol{\theta}^{[f]})=\mathrm{vec}\left(\mathbf{J}^{[f]}_{i,p}(\boldsymbol{\theta}^{[f]})\mathbf{C}^{[f]}_{i}
\mathbf{J}_{i,q}^{[f]^H}(\boldsymbol{\theta}^{[f]})\right)
=\left(\mathbf{J}^{[f]^{\ast}}_{i,q}(\boldsymbol{\theta}^{[f]})\otimes
\mathbf{J}^{[f]}_{i,p}(\boldsymbol{\theta}^{[f]})\right)\mathbf{c}^{[f]}_{i}$ and $\mathbf{c}^{[f]}_{i}=\mathrm{vec}(\mathbf{C}^{[f]}_{i})$. If we note $B=\frac{M(M-1)}{2}$ the global number of antenna pairs, all cross-correlation measurements are given by the following $4B \times 1$ full observation vector at frequency $f$
\begin{equation}
\label{express_x}
\mathbf{x}^{[f]}=\left[\mathbf{x}^{[f]^{T}}_{12},\mathbf{x}^{[f]^{T}}_{13},
\ldots,
\mathbf{x}^{[f]^{T}}_{(M-1)M}\right]^{T}.
\end{equation}
%=\sum_{i=1}^{D}\left[\mathbf{s}^{[f]^T}_{i,12}(\boldsymbol{\theta}^{[f]}),
%\ldots,
%\mathbf{s}^{[f]^T}_{i,(M-1)M}(\boldsymbol{\theta}^{[f]})\right]^{T}+\left[\mathbf{n}^{[f]^T}_{12},
%\ldots,
%\mathbf{n}^{[f]^T}_{(M-1)M}\right]^{T}.

Each Jones matrix $\mathbf{J}^{[f]}_{i,p}(\boldsymbol{\theta}^{[f]})$ in (\ref{start}) can be decomposed into a particular sequence of individual Jones terms \cite{smirnov2011revisiting}. Each of them corresponds to a particular effect along the signal propagation path. More specifically, we can write 
\begin{equation}
 \label{model_regime3}
\mathbf{J}^{[f]}_{i,p}(\boldsymbol{\theta}^{[f]})=\mathbf{G}^{[f]}_p(\mathbf{g}^{[f]}_p)\mathbf{H}^{[f]}_{i,p}\mathbf{Z}^{[f]}_{i,p}(\varphi^{[f]}_{i,p})\mathbf{F}^{[f]}_{i,p}(\vartheta^{[f]}_{i,p})
\end{equation}
where $\mathbf{G}^{[f]}_p(\mathbf{g}^{[f]}_p)=\mathrm{diag}\{\mathbf{g}^{[f]}_p\}$ stands for the complex electronic gain matrix and $\mathbf{H}^{[f]}_{i,p}$ is assumed known, including the geometric delay and beam pattern effects \cite{thompson2008interferometry}. Due to large distances between antennas and wide field-of-view, propagation through the ionosphere leads to a per-antenna direction dependent phase delay $\mathbf{Z}^{[f]}_{i,p}(\varphi^{[f]}_{i,p})=\exp\Big(j\varphi^{[f]}_{i,p}\Big)\mathbf{I}_2$ but also to a rotation effect \cite{davies1990ionospheric}, named Faraday rotation and written as 
%\begin{equation}
%\label{form_F}
$\mathbf{F}^{[f]}_{i,p}(\vartheta^{[f]}_{i,p}) = \begin{bmatrix}
    \cos(\vartheta^{[f]}_{i,p}) & -\sin(\vartheta^{[f]}_{i,p}) \\
   \sin(\vartheta^{[f]}_{i,p
   }) & \cos(\vartheta^{[f]}_{i,p})
\end{bmatrix}.$
%\end{equation}
Thus, the unknown parameter vector of interest which describes all Jones matrices at frequency $f$ reads 
%$ \boldsymbol{\theta}^{[f]}=  [\vartheta^{[f]}_{1,1},\hdots, \vartheta^{[f]}_{D,M}, \exp(j\varphi^{[f]}_{1,1}),\hdots,
%\exp(j\varphi^{[f]}_{D,M}), \mathbf{g}_1^{[f]^T},\hdots, \\ \mathbf{g}_M^{[f]^T}]^T.$
%\begin{align}
%\nonumber
$ \boldsymbol{\theta}^{[f]}=% & 
[\vartheta^{[f]}_{1,1},\hdots, \vartheta^{[f]}_{D,M},\exp(j\varphi^{[f]}_{1,1}),\hdots, %\\ & 
\exp(j\varphi^{[f]}_{D,M}),\mathbf{g}_1^{[f]^T},\hdots, \\ \mathbf{g}_M^{[f]^T}]^T.$
%\end{align}
%in which
%\begin{equation}
%\boldsymbol{\epsilon}^{[f]}=[\vartheta^{[f]}_{1,1},\hdots, \vartheta^{[f]}_{D,M},\exp(j\varphi^{[f]}_{1,1}),\hdots, 
%\exp(j\varphi^{[f]}_{D,M})]^T.
%\end{equation}

\subsection{Robust noise modeling}

In order to propose a robust calibration algorithm, we study the flexible CG representation which encompasses a broad range of heavy-tailed distributions and can be written as \cite{yao2003spherically,wang2006maximum}
\begin{equation}
\label{sirp}
\mathbf{n}^{[f]}_{pq} = \sqrt{\tau^{[f]}_{pq}} \  \boldsymbol{\mu}^{[f]}_{pq},
\end{equation} 
where $\tau^{[f]}_{pq}$ denotes the positive texture random variable and $\boldsymbol{\mu}^{[f]}_{pq} \sim \mathcal{CN}(\mathbf{0},\boldsymbol{\Omega}^{[f]})$ with a uniform prior on the positive semidefinite cone for $\boldsymbol{\Omega}^{[f]}$. To avoid any ambiguity \cite{zhang2016mimo,ollierjournal}, we impose an arbitrary constraint on the speckle covariance matrix $\boldsymbol{\Omega}^{[f]}$, \textit{e.g.}, $\mathrm{tr}\left\{\boldsymbol{\Omega}^{[f]} \right\}=1$. Therefore, in our context, calibration requires estimating not only parameters of interest $\boldsymbol{\theta}^{[f]}$ but also noise terms $\boldsymbol{\tau}^{[f]}=[\tau^{[f]}_{12},\tau^{[f]}_{13},\ldots,\tau^{[f]}_{(M-1)M}]^{T}$ and $\boldsymbol{\Omega}^{[f]}$.

%In LaTeX, to start a new column (but not a new page) and help balance the last-page column lengths, you can use the command ``$\backslash$pagebreak'', as demonstrated near the end of the LaTeX source for this template.

\section{MAXIMUM A POSTERIORI ESTIMATION}
\label{sec:MLestim}

\subsection{Principle of the proposed algorithm}

In this section, we investigate robust calibration from a Bayesian perspective. For this purpose, we focus on the texture realizations and derive MAP-based estimators.
%based on the probability density function (pdf) of $\tau^{[f]}_{pq}$, . 

Assuming independence of $\mathbf{n}^{[f]}_{pq}$ \textit{w.r.t.} frequency channels $f$ and antenna pairs $(p,q)$, the conditional log-likelihood function is written as
\begin{align}
\label{likelihood}
\nonumber
 &  \mathcal{L}_C = \ln p \left(\{\mathbf{x}^{[f]}\}_{f \in \mathcal{F}} \ | \ \{\boldsymbol{\tau}^{[f]};\boldsymbol{\theta}^{[f]},\boldsymbol{\Omega}^{[f]}\}_{f \in \mathcal{F}}\right)  =
  -\sum_{f \in \mathcal{F}} \\ &    \sum\limits_{pq} 
\left(\frac{1}{\tau^{[f]}_{pq}}\mathbf{u}^{[f]^{H}}_{pq}(\boldsymbol{\theta}^{[f]})
\boldsymbol{\Omega}^{[f]^{-1}}\mathbf{u}^{[f]}_{pq}(\boldsymbol{\theta}^{[f]})+
\ln | \pi
\tau^{[f]}_{pq} \boldsymbol{\Omega}^{[f]} |\right)
\end{align}
in which $\mathbf{u}^{[f]}_{pq}(\boldsymbol{\theta}^{[f]}) = \mathbf{x}^{[f]}_{pq}- \sum\limits_{i=1}^{D}\mathbf{s}^{[f]}_{i,pq}(\boldsymbol{\theta}^{[f]})$. 
Thanks to Bayes' law, the joint log-likelihood function, denoted as $\mathcal{L}_J$, reads
\begin{equation}
\label{likelihood_joint}
 \mathcal{L}_J=\mathcal{L}_C + 
 \sum_{f \in \mathcal{F}}\sum\limits_{pq}
 \ln \left(  
 p (\tau^{[f]}_{pq}; \boldsymbol{\varphi}^{[f]})  
\right)
\end{equation}
in which $\boldsymbol{\varphi}^{[f]}$ stand for the corresponding unknown hyperparameters which need to be estimated. 

The MAP estimator consists in maximizing (\ref{likelihood_joint})  \textit{w.r.t.} each unknown individual parameter. The expressions being mutually dependent, we adopt an iterative procedure leading to the IMAPE. More specifically, we optimize alternatingly \textit{w.r.t.} $\boldsymbol{\theta}^{[f]}$, $\boldsymbol{\tau}^{[f]}$, $\boldsymbol{\varphi}^{[f]}$  and $\boldsymbol{\Omega}^{[f]}$ in a step-wise approach. At each step, the remaining parameters are assumed fixed and known from the previous iteration. The overall procedure of the proposed IMAPE is shown in Table 1 hereafter.
%global pattern of the algorithm

Estimation of $\{\hat{\boldsymbol{\theta}}^{[f]}\}_{f \in \mathcal{F}}$ in step 1 of the IMAPE, for a given $\{\boldsymbol{\tau}^{[f]}\}_{f \in \mathcal{F}}$ and $\{\boldsymbol{\Omega}^{[f]}\}_{f \in \mathcal{F}}$,  is detailed in \cite{ollier2018,ollier2018article}. These parameters of interest vary specifically with frequency such that we can reformulate the problem as a constrained optimization scheme and solve it distributedly with a network of agents and the consensus ADMM procedure \cite{bertsekas1989parallel}. Let us note that $\partial \mathcal{L}_J = \partial \mathcal{L}_C$ if we take the derivative \textit{w.r.t.} $\boldsymbol{\theta}^{[f]}$ or $\boldsymbol{\Omega}^{[f]}$ and as mentionned previously, the normalization step $\hat{\boldsymbol{\Omega}}^{[f]}=\frac{\hat{\boldsymbol{\Omega}}^{[f]}}{\mathrm{tr}\left\{\hat{\boldsymbol{\Omega}}^{[f]}\right\}}$ is added in step 3 of the IMAPE for proper identifiability. Consequently, in what follows, we derive and study for each prior distribution $ p (\tau^{[f]}_{pq}; \boldsymbol{\varphi}^{[f]})$ the corresponding IMAPE  and highlight how the estimation of Jones matrices' parameters is affected in the numerical simulations section. Thus, we focus on steps 2 to 4 of the IMAPE. Furthermore, since the reasoning is equivalent for each frequency $f$, we omit the frequency dependence in the rest of the paper for sake of clarity.

\LinesNotNumbered
%[t!]
\begin{algorithm}[t!]
\SetAlgorithmName{Table 1}{}{} \caption{\textbf{IMAPE} Iterative MAP estimator}
\SetKwInOut{input}{input} \SetKwInOut{output}{output}
\SetKwInOut{initialize}{initialize}
\input{$D$, $M$, $B$, $\{\{\mathbf{C}^{[f]}_{i}\}_{i=1,\hdots,D},\mathbf{x}^{[f]}\}_{f \in \mathcal{F}}$}
\output{$\{\hat{\boldsymbol{\theta}}^{[f]}\}_{f \in \mathcal{F}}$}
\initialize{$\{\hat{ \boldsymbol{\Omega}}^{[f]} \leftarrow \boldsymbol{\Omega}^{[f]}_{\mathrm{init}}\}_{f \in \mathcal{F}}$, \\ $\{\hat{\tau}^{[f]}_{pq} \leftarrow \tau^{[f]}_{pq_{\mathrm{init}}}\}_{f \in \mathcal{F},  \ p<q, \ (p,q) \in \{1,\hdots,M\}^2}$}
\While
{stop criterion unreached}
{
\nextnr$\{\hat{\boldsymbol{\theta }}^{[f]}\}_{f \in \mathcal{F}} =\argmax\limits_{\{\boldsymbol{\theta }^{[f]}\}_{f \in \mathcal{F}} }
\left\{\mathcal{L}_C\right\}$, see \cite{ollier2018,ollier2018article}\\
\nextnr  Estimation of $\boldsymbol{\varphi}^{[f]}$ for $f \in \mathcal{F}$\\ 
\nextnr
$\hat{\boldsymbol{\Omega}}^{[f]}= \frac{1}{B}
\sum\limits_{pq} \frac{\mathbf{u}^{[f]}_{pq}(\hat{\boldsymbol{\theta}}^{[f]})
\mathbf{u}^{[f]^{H}}_{pq}(\hat{\boldsymbol{\theta}}^{[f]})}{\hat{\tau}^{[f]}_{pq}}$ \\  $\hat{\boldsymbol{\Omega}}^{[f]}=\frac{\hat{\boldsymbol{\Omega}}^{[f]}}{\mathrm{tr}\left\{\hat{\boldsymbol{\Omega}}^{[f]}\right\}}$ for $f \in \mathcal{F}$
\\
\nextnr  Estimation of $\boldsymbol{\tau}^{[f]}$ for $f \in \mathcal{F}$\\
}
\end{algorithm}
%\argmin\limits_{\boldsymbol{\theta } }
%\left\{\sum\limits_{pq}\frac{1}{\hat{\tau}_{pq}}\mathbf{a}_{pq}^{H}(\boldsymbol{\theta})
%\hat{\boldsymbol{\Omega}}^{-1}\mathbf{a}_{pq}(\boldsymbol{\theta})\right\}

\subsection{Study of different texture priors}

In the Bayesian approach, the expression of the joint log-likelihood in (\ref{likelihood_joint}) depends on $ p (\tau_{pq}; \boldsymbol{\varphi})$, which describes the statistics of the texture variable. In this section, we study different kinds of heavy-tailed CG distributions obtained with different texture priors, available in closed-form \cite{ollila2012complex,jay2002detection}. Then, the corresponding estimators for $\boldsymbol{\varphi}$, $\boldsymbol{\Omega}$ and $\boldsymbol{\tau}$ are deduced. 

\subsubsection{K-distribution}

The noise vector $\mathbf{n}_{pq}$ is said to follow the K-distribution \cite{ollila2012complex} if the texture $\tau_{pq}$ is generated as
\begin{equation}
\label{pdf_K}
p(\tau_{pq};a,b)=\frac{1}{\Gamma(a)b^a} \tau_{pq}^{a-1} \exp\left(-\frac{\tau_{pq}}{b}\right)
\end{equation}
which is a gamma distribution characterized by the shape parameter $a$ and the scale parameter $b$, such that, $\boldsymbol{\varphi}=[a,b]$.
Inserting (\ref{pdf_K}) into (\ref{likelihood_joint}) leads to 
\begin{align}
\nonumber
 \mathcal{L}_J= & \mathcal{L}_C + 
(a-1)\sum_{pq} \ln(\tau_{pq}) - \frac{ \sum\limits_{pq} \tau_{pq}}{b} \\ & -B \ln(\Gamma(a))-B a \ln(b).
\end{align}

Solving $\partial \mathcal{L}_J / \partial \tau_{pq}=0$ gives the following estimate for the texture
% \overset{+}{-}
\begin{equation}
\label{estim_tau}
\hat{\tau}_{pq}=\frac{(a-5)b + \Big((a-5)^2 b^2+4 b \mathbf{u}^{^H}_{pq}(\boldsymbol{\theta})\boldsymbol{\Omega}^{-1}\mathbf{u}_{pq}(\boldsymbol{\theta})\Big)^{1/2}}{2}.
\end{equation}

From $\partial \mathcal{L}_J / \partial \boldsymbol{\varphi}=0$, we deduce the following analytical expression for the scale
\begin{equation}
\label{estim_b}
\hat{b}=\frac{\sum\limits_{pq} \tau_{pq} } { B a}.
\end{equation}
Finally, the shape parameter has to satisfy the following equation
\begin{equation}
\label{estim_a}
-B \Psi(\hat{a})+\sum_{pq} \ln (\tau_{pq}) -B \ln(b)=0
\end{equation}
from which the shape parameter can be easily calculated numerically.

For the K-distributed noise model, steps 2 to 4 of the IMAPE are given in Table 2. Let us note that (\ref{estim_b}) was substituted into (\ref{estim_a}) and likewise, (\ref{estim_tau}) was inserted into the expression of $\hat{\boldsymbol{\Omega}}$ due to our iterative approach.

In the following sections, using the same technique, we only give the results for the estimates of the unknown noise parameters and deduce similar procedures as in Table 2.

\begin{algorithm} [t]
\SetAlgorithmName{Table 2}{}{} \caption{\textbf{IMAPE} K-distributed noise}
\SetKwInOut{input}{input} \SetKwInOut{output}{output}
\SetKwInOut{initialize}{initialize}
\setcounter{AlgoLine}{1}
\nextnr  Obtain $\hat{a}$ solving  $-B \Psi(a)+\sum_{pq} \ln(\hat{\tau}_{pq}) -B \ln(\frac{\sum_{pq} \hat{\tau}_{pq} } { B a})=0$\\
\setcounter{AlgoLine}{2}
Obtain $\hat{b}$ with (\ref{estim_b})
\\
\nextnr  Obtain $\hat{\boldsymbol{\Omega}}$ with 
$\hat{\boldsymbol{\Omega}}= \frac{2}{B}
\sum\limits_{pq} \frac{\mathbf{u}_{pq}(\hat{\boldsymbol{\theta}})
\mathbf{u}^{H}_{pq}(\hat{\boldsymbol{\theta}})}{(\hat{a}-5)\hat{b} + \Big((\hat{a}-5)^2 \hat{b}^2+4 \hat{b} \mathbf{u}^{^H}_{pq}(\hat{\boldsymbol{\theta}})\hat{\boldsymbol{\Omega}}^{^{-1}}\mathbf{u}_{pq}(\hat{\boldsymbol{\theta}})\Big)^{1/2}}$ $\hat{\boldsymbol{\Omega}}=\frac{\hat{\boldsymbol{\Omega}}}{\mathrm{tr}\left\{\hat{\boldsymbol{\Omega}}\right\}}$\\
\nextnr Obtain $\hat{\boldsymbol{\tau}}$ with (\ref{estim_tau})
\\
\end{algorithm}

\subsubsection{Student's t distribution}
\label{sub_student}

If the texture parameter follows an inverse gamma function, given by
\begin{equation}
\label{pdf_T}
p(\tau_{pq};a,b)=\frac{b^a}{\Gamma(a)} \tau_{pq}^{-a-1} \exp\left(-\frac{b}{\tau_{pq}}\right),
\end{equation}
 the corresponding noise model is a Student's t distribution \cite{lange1989robust}. With (\ref{pdf_T}), (\ref{likelihood_joint}) and the derivative \textit{w.r.t.} $\tau_{pq}$, we obtain
\begin{equation}
\label{estim_tauT}
\hat{\tau}_{pq}=\frac{b+ \mathbf{u}^{^H}_{pq}(\boldsymbol{\theta})\boldsymbol{\Omega}^{-1}\mathbf{u}_{pq}(\boldsymbol{\theta})}{a+5}
\end{equation}
while hyperparameters can be deducted from
%\begin{equation}
%\label{estim_bT}
$\hat{b}=\frac{B a } { \sum\limits_{pq} \frac{1}{\tau_{pq}}}$
%\end{equation}
and 
\begin{equation}
\label{estim_aT}
-B \Psi(a)-\sum_{pq} \ln( \tau_{pq})+B \ln(b)=0.
\end{equation}

From now on, we introduce texture prior distributions with only one hyperparameter to estimate.

\subsubsection{Cauchy distribution}

Using the inverse gamma prior distribution (\ref{pdf_T}) with fixed $a=1$ leads to Cauchy noise modeling for $\mathbf{n}_{pq}$ \cite{jay2002detection}. This is a particular case of section \ref{sub_student} where (\ref{estim_aT}) becomes $\hat{a}=1$.

\subsubsection{Laplace distribution}

Let us consider the following exponential distribution as prior:
%\begin{equation}
%\label{pdf_L}
$p(\tau_{pq};\lambda) = \lambda \exp\left(-\lambda \tau_{pq}\right)$
%\end{equation}
characterized by rate parameter $\lambda$, and corresponding to Laplace noise \cite{jay2002detection}. After some calculus, the texture estimate reads  \\
\begin{equation}
\label{estim_tauL}
\hat{\tau}_{pq}=\frac{-4 + \Big(16+4 \lambda \mathbf{u}^{^H}_{pq}(\boldsymbol{\theta})\boldsymbol{\Omega}^{^{-1}}\mathbf{u}_{pq}(\boldsymbol{\theta})\Big)^{1/2}}{2 \lambda}
\end{equation}
while the single parameter $\lambda$ is given by
%\begin{equation}
%\label{estim_lambdaL}
$\hat{\lambda}=\frac{B}{\sum\limits_{pq} \tau_{pq}}.$
%\end{equation}

\subsubsection{Inverse Gaussian compound-Gaussian (IG-CG) distribution}

Finally, we consider the case when the texture variable is modeled by an inverse Gaussian distribution, also known as the Wald distribution, with shape $\lambda$ and assumed unit mean \cite{ollila2012complex}. Thus, its probability density function is of the form
%\begin{equation}
$p(\tau_{pq};\lambda)=\Big(\frac{\lambda}{2 \pi}\Big)^{1/2} \tau_{pq}^{-3/2} \exp\left(\frac{-\lambda (\tau_{pq}-1)^2}{2 \tau_{pq}}\right).$
%\end{equation}

The resulting CG is called an IG-CG distribution and the corresponding noise parameters are given by  \\
\begin{equation}
\hat{\tau}_{pq}=\frac{-11 +(121+4 \lambda (\lambda +2\mathbf{u}^{^H}_{pq}(\boldsymbol{\theta})\boldsymbol{\Omega}^{^{-1}}\mathbf{u}_{pq}(\boldsymbol{\theta})))^{1/2}}{2 \lambda}
\end{equation}
and
%\begin{equation}
$\hat{\lambda}=\frac{B}{\sum\limits_{pq} \frac{(\tau_{pq}-1)^2}{\tau_{pq}}}.$
%\end{equation}

\section{NUMERICAL SIMULATIONS}
\label{sec:Simus}

\begin{figure}[t!] %%%% ou htb
  \centering
  \centerline{\includegraphics[height=6cm,width=8.5cm]{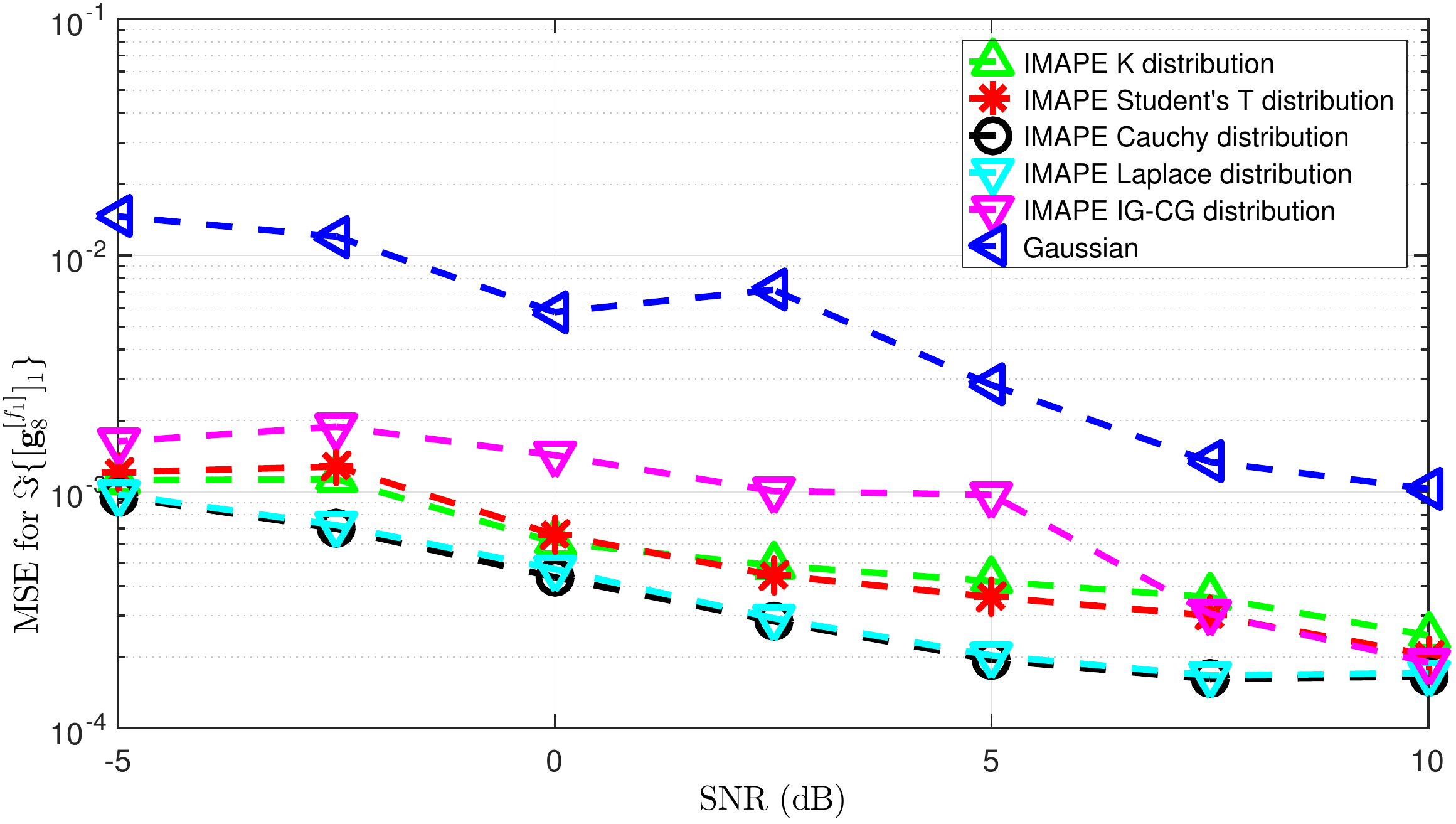}}
\caption{Evolution of the MSE for the imaginary part of a given complex gain  as a function of the SNR for different kinds of CG distributions.} \label{fig:res1}
\end{figure}

\begin{figure}[t!] %%%% ou htb
  \centering
  \centerline{\includegraphics[height=6cm,width=8.5cm]{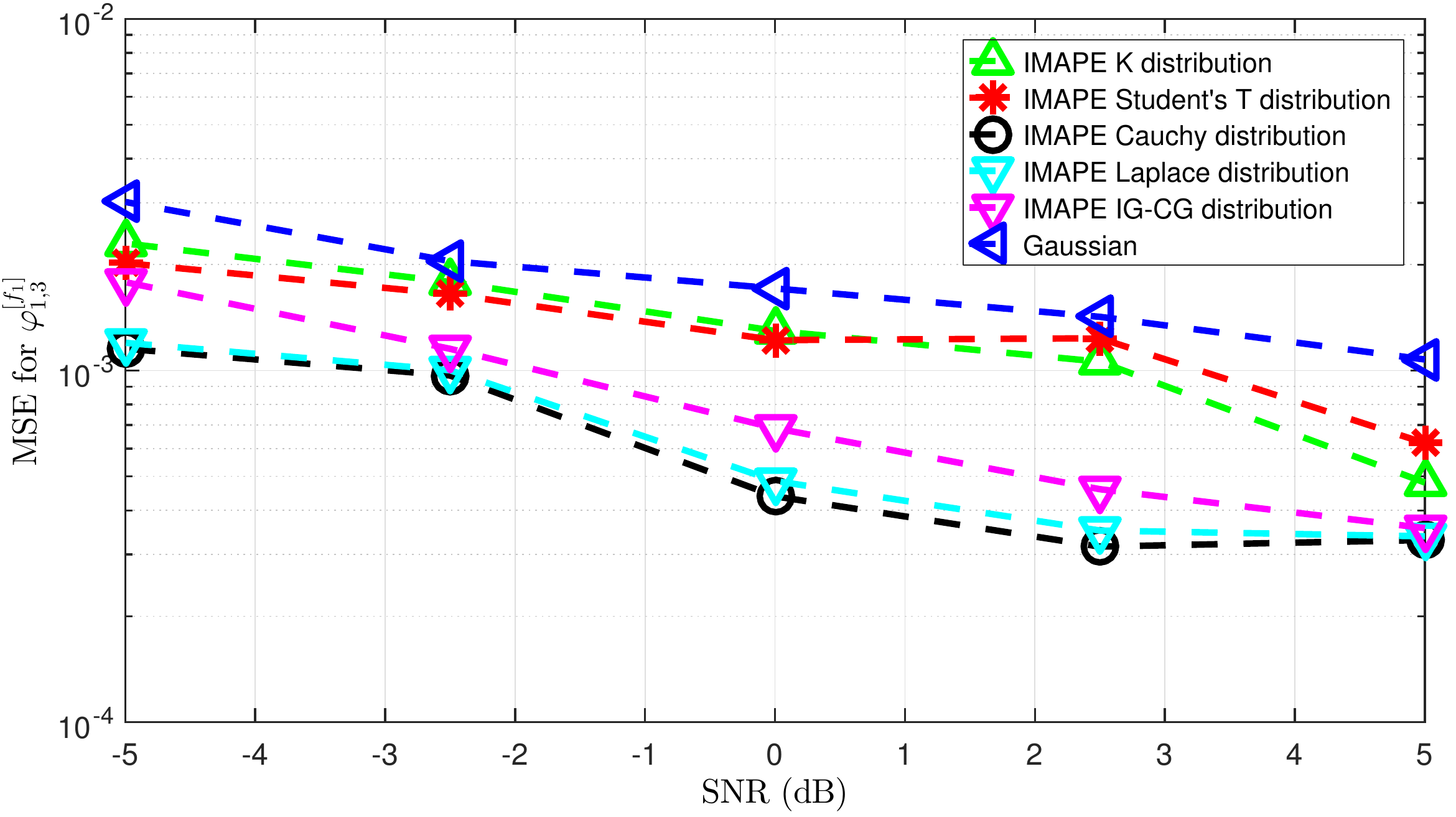}}
\caption{Evolution of the MSE for a given ionospheric phase delay  as a function of the SNR for different kinds of CG distributions.} \label{fig:res2}
\end{figure}

%%%%kinds of CG distributions

%%%

We aim to compare the estimation performance of the IMAPE for selected examples of CG-based noise models, in a realistic situation for radio astronomy. Let us consider $M=8$ antennas in a 2-dimensional sensor array, $D=2$ bright calibrator sources and $D'=4$ weak background sources (at least 10 times weaker than the calibrator sources) which are spatially randomly distributed following a uniform distribution. Perturbation effects are randomly generated thanks to the measurement equation in (\ref{model_regime3}) and white Gaussian background noise is added to the data. Due to the presence of faint unknown sources, the noise can no longer be considered Gaussian. To highlight the robustness of the proposed Bayesian method, we compare with traditional calibration based on the minimization of a least squares cost function, \textit{i.e.}, by assuming independent Gaussian noise with no Bayesian approach.
%\cite{yatawatta2009radio}

In Figures \ref{fig:res1} and \ref{fig:res2}, we plot the mean square error (MSE) of the imaginary part of one given complex gain and one ionospheric phase delay, respectively, as a function of the signal-to-noise ratio (SNR) for different noise modeling. The SNR is defined as the ratio of the normalized power of the $D$ calibrator sources over the sum of normalized power of the $D'$ background sources and a noise factor.
The commonly used Gaussian noise assumption has poor performance as expected while we notice some dissimilarities for the IMAPE depending on which $  p (\tau_{pq}; \boldsymbol{\varphi})$ is considered. We note that the IMAPE based on Cauchy distribution provides the best performance which can be selected as a competitive noise model for Bayesian calibration of radio interferometers.

%reveals to be the most accurate to model a non-Gaussian process in our application.
% for our radio interferometric measurements. 

\section{CONCLUSION}
\label{sec:Conclusion}

In this paper, we adopted a Bayesian approach for robust calibration of radio interferometers where robustness needs to be ensured due to the presence of outliers. To address this problem, we derived specific MAP-based estimators which exploit the statistical distribution of the texture in the CG representation. Such family covers a wide variety of non-Gaussian models, obtained with different prior distributions. Application to simulated data shows that the choice of the prior actually affects the estimation of the perturbation effects and reveals to be more or less accurate to model the noise contribution in our application. This leads us to select the Cauchy distribution as a suitable model for the noise. 
Additional comparison with realistic data simulations in the software Meqtrees and real data will be performed in future work.

\begin{small}
\bibliographystyle{IEEEbib}
%\bibliography{nab}

%\bibliographystyle{IEEEtran}
\bibliography{biblio}
%\bibliography{strings,refs}

\end{small}

\end{document}